\documentclass[8pt,aps,prb,nofootinbib,showpacs,twocolumn,latexsym]{revtex4}
\usepackage{amssymb}

\usepackage{latexsym}
\usepackage[dvips]{graphicx}
\usepackage[dvips]{color}
\usepackage{amsmath}
\usepackage[mathcal]{eucal}

\begin{document}

\title{Temperature dependence of spin susceptibility in
two-dimensional~Fermi~liquid systems.}
\author{ A.~Shekhter, and A.M.~Finkel'stein}

\begin{abstract}
We consider the non-analytic terms in the spin susceptibility arising as a
result of rescaterring of pairs of quasiparticles. We emphasize the
importance of rescattering in the Cooper channel for the analysis of the
temperature dependences in the two-dimensional electron systems in the
ballistic regime. In the calculation of the linear in $T$ term we use
angular harmonics in the Cooper channel, because for each harmonic the
interaction amplitude is renormalized independently. We observe, that as a
consequence of strong renormalizations in the Cooper ladder, the temperature
derivative of the spin susceptibility may change its sign at low
temperatures.
\end{abstract}

\affiliation{Department of Condensed Matter Physics, the Weizmann Institute of Science,
Rehovot, 76100, Israel and \\
Argonne National Laboratory, Materials Science Division, Argonne, IL 60439}
\pacs{71.27.+a,75.40.Cx,71.10.Pm}
\maketitle

\renewcommand{\Re}{\mathrm{Re}\,} \renewcommand{\Im}{\mathrm{Im}\,}

\section{Introduction.}

Linear in temperature corrections in the spin susceptibility of the
two-dimensional (2D) electron gas has been discussed intensively in the past
decade.\cite%
{Baranov1993,Belitz1996b,Takahashi1998,Misawa1999,ChitovLett2001,Chubukov2003,Galitski2005,GlazmanB2005,Betouras2005}
Our interest to this question is motivated by a recent observation of a
strong temperature dependence in the spin susceptibility in the silicon
metal-oxide-semiconductor field-effect transistor (Si-MOSFET).\cite{Prus2003}
It has been noticed by Misawa\cite{Misawa1999} that linear in $T$
corrections, if exist, should be a result of a non-analytic behavior of the
thermodynamic potential at a small temperature and magnetic field. Indeed, a
naive Taylor expansion term $H^{2}T$ in the thermodynamic potential
corresponding to the linear in $T$ correction to the spin susceptibility
would violate the third Law of thermodynamics. Therefore, there should be a
strong dependence on the order of limits $H\rightarrow 0,T\rightarrow 0$ in
the thermodynamic potential. Obviously, to analyze such a non-analytic
behavior one has to go beyond the standard theory of the Fermi liquid
systems. Still, we use a machinery of the microscopic Fermi liquid theory as
our starting point.

It is known that a pair of Green's functions has a singular behavior when it
has a momentum transfer close to $2p_{F}$. In the polarization operator this
singularity reveals itself as the Kohn anomaly at $2p_{F}$. In the case of
2D, the Kohn anomaly leads to a cusp-like dependence on the transferred
momentum and frequency.\cite{Stern1980,Gold1986,DasSarma1986} This cusp-like
behavior is related to the fact that for a $2p_{F}$ momentum transfer the
two patches on the Fermi surface are located on the opposite sides and are
parallel to each other. The cusp-edges are sensitive to a relative shift of
the chemical potentials. When the Fermi surface is spin-split due to
magnetic field, the sharp effect of shifting of the two cusps near the point
when they touch generates a non-analytic temperature dependencies in the
thermodynamics quantities. This touching of singularities gives a clue for
understanding why there is an anomaly in the spin-susceptibility. The
analysis of the temperature correction with the use of the non-analytic
parts in the product of two pairs of Green's functions discussed in this
paper is different from the previous calculations; e.g. Ref.~[%
\onlinecite{Betouras2005}]. We believe that it makes the origin of the
non-analytical behavior of the thermodynamic potential more transparent. The
dependence on the order of limits $H,T\rightarrow 0$ are explicitly
demonstrated in this calculation.

\begin{figure}[tbp]
\centerline{\includegraphics[width=0.45%
\textwidth]{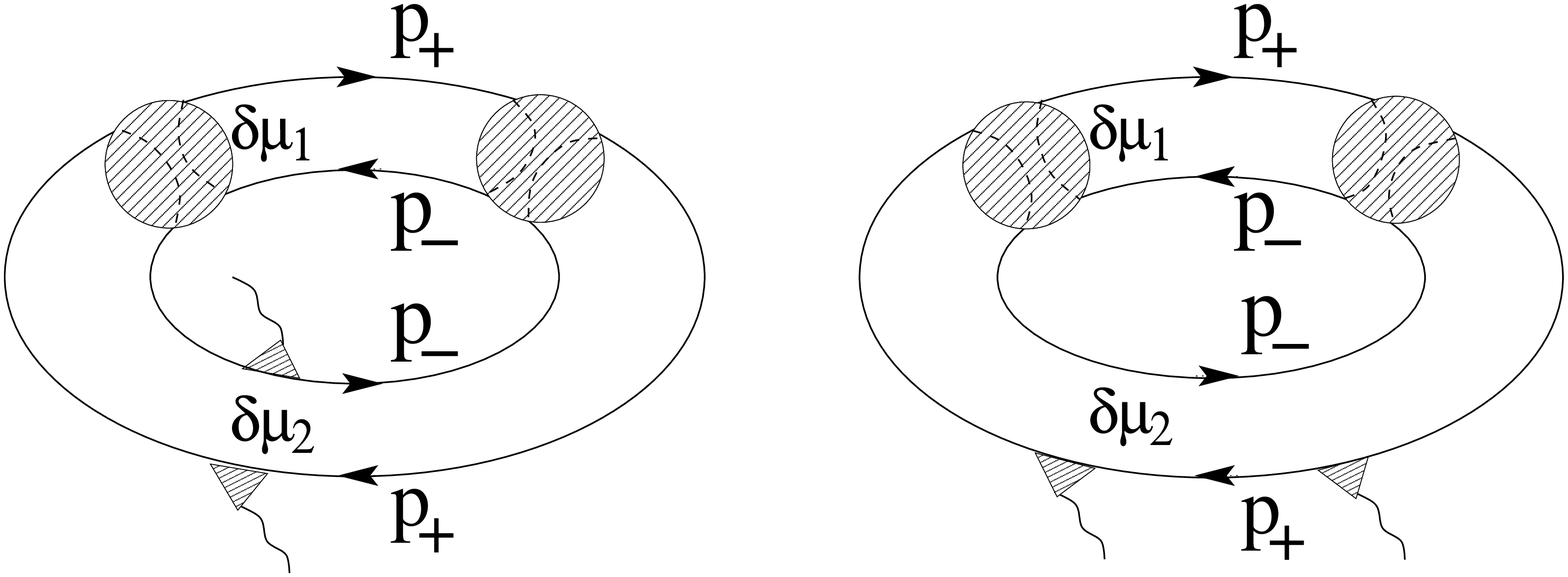}}
\caption{ The diagrams with four Green's functions in the thermodynamic
potential which produce the anomalous term in the spin susceptibility. $%
\mathbf{p}_{+}$ and $\mathbf{p}_{-}$ indicate the directions of the momenta
of electrons: $\mathbf{p}_{+}\approx -\mathbf{p}_{-}$. The chemical
potential shifts $\protect\delta \protect\mu _{1}$ and $\protect\delta 
\protect\mu _{2}$ has been introduced in each of the two pairs of Green's
function. The spin-susceptibility can be generated from these diagrams by
taking the second derivative with respect to $\protect\delta\protect\mu%
_{1,2} $. The triangles attached to the external vertices indicate the
Fermi-liquid renormalization of the static spin vertices.}
\label{fig:4Greenfunc}
\end{figure}

The calculation of the non-analytic term in the thermodynamic potential
containing two pairs of Green's functions shows that the anomalous
temperature dependence originates from small frequencies and momenta close
to the Fermi surface. The momenta are restricted to the narrow angular
intervals corresponding to scattering with a $2p_{F}$ momentum transfer as
it is shown in Fig.~\ref{fig:4Greenfunc}. The amplitude associated with such
scattering processes is natural to be called the backward scattering
amplitude and will be denoted as $\Gamma(\pi)$. Because of a special
configuration of its four momenta, the amplitude $\Gamma(\pi)$ can act also
within the Cooper channel.\cite{zerosound} Moreover, the product of two $2p_{F}$-pairs of Green's functions by an appropriate twisting of the GreenÕs functions can be read as the product of two pairs
in the Cooper channel (or two particle-hole pairs in the zero-sound channel~\cite{PNAS2006}).
In this paper we demonstrate that the non-analytic terms, associated
previously only with $2p_{F}$-scattering, should be analyzed having in mind
the rescattering of pairs in the Cooper channel.\cite{third} We emphasize
here the rescattering in the Cooper channel as the only source of the
temperature-dependent renormalizations of the thermodynamic quantities of
the Fermi liquid systems in the ballistic regime. The analysis of the
renormalization of the anomalous temperature dependencies is performed in
terms of angular harmonics in the Cooper channel, because each harmonic is
renormalized differently. In the calculation of the linear in $T$ term with
the use of harmonics in the Cooper channel we have not assumed a priory the
special importance of the backward scattering amplitude $\Gamma(\pi)$.
Interestingly, we obtain this fact as a result of independent calculation of
the product of two pairs.

We show that in the Cooper channel ladder the non-analytic terms in the
spin-susceptibility, $\delta\chi $, are generated by the products with
arbitrary number of the pairs of Green's functions, i.e., not only a product
of the two pairs. This happens because a correlation function describing the
propagation of a pair of quasiparticles in the Cooper channel has a dynamic
part which depends on the ratio $\omega/qv_{F}$. (In the microscopic Fermi
liquid theory\cite{Pitaevskii} the scattering amplitude in the zero sound
channel has exactly the same feature.) Consequently, we obtain that $%
\partial\delta\chi/\partial{T}$ is given by a power series in the
renormalized Cooper channel amplitudes that does not reduce to the
renormalization group generalization of the result obtained in the second
order. The truncation of this series is possible only when the Cooper
channel amplitudes are small.

This paper is organized as follows. In Section~2 a general discussion of the
anomalous temperature dependences in the Fermi liquid systems is presented.
We start by stressing that to get anomalous temperature dependencies it is
necessary to pinpoint non-analyticities in the thermodynamic potential to
prevent regular Sommerfeld's expansion at low temperatures. To get a clue of
the origin of the non-analytic terms, an analogy with an auxiliary model
system is presented. In the end of this Section we explain why in the
analysis of the non-analytic terms in the thermodynamic potential the ladder
diagrams are of particular importance. In Section~3 the physical
consequences of the renormalizations in the Cooper channel are discussed in
connection with the temperature dependence in the spin susceptibility
observed in the Si-MOSFET.\cite{Prus2003} It is argued that at low
temperature when the interaction amplitudes are strongly renormalized their
own temperature dependence may overcome the linear in temperature factor.
This may be a possible explanation of the observed sign of the temperature
dependence of the spin susceptibility in the 2D electron gas. All technical
details of the calculations are moved in Appendices. We first present
details of the calculations of the linear in $T$ term in the spin
susceptibility originating from two $2p_{F}$-pairs of quasiparticles. In
Appendix~\ref{sec:2pf} we obtain an expression which demonstrate explicitly
how the dependence on the order of taking the limits $H\rightarrow
0,T\rightarrow 0$ appears in the thermodynamic potential (for more
discussions on this subject see also Ref.~[\onlinecite{PNAS2006}]).
Next, in Appendices~\ref{sec:Cooperchannel} and~\ref{sec:hnm} the
temperature corrections are analyzed within the Cooper channel. In the
concluding Section~4 we give several historical remarks. In particular we
stress a rather unique nature of the temperature dependences in 2D.

\section{General discussion of anomalous terms in $\protect\chi$.}

Let us discuss the origin of the anomalous temperature dependences. The
calculation of interaction corrections to a thermodynamic quantity requires
a summation over bosonic frequency $i\omega _{n}$. Transforming the sum into
an integral along the real frequency axis results in an integral of the form 
$\int{d}\omega\coth(\beta\omega/2)f(\omega)$. If $f(\omega)$ is smooth and
regular in the vicinity of $\omega =0$, the standard Sommerfeld expansion
will involve only even powers of temperature. In order to get a term with an
odd power of temperature (e.g., linear in $T$ correction to the
spin-susceptibility), $f$ should be non-analytic preventing the Taylor
expansion at $\omega=0$. Let us see how such non-analyticity develops in the
case of the 2D electron gas. It is shown in Appendices~\ref{sec:2pf} and~\ref%
{sec:hnm} that both a $2p_{F}$-pair of Green's functions and a pair in the
Cooper channel have non-analytic parts (edge parts). Together these edge
parts act as a sort of an ''edge-mode'' propagating in the opposite
directions. In the case of a $2p_{F}$-pair they are denoted as $\Pi _{r}$
and $\Pi _{l}$, i.e., ''right'' and ''left'', reflecting the combinations $%
\omega \pm {v}_{F}\delta {q}$ in the denominators of Eq.~(\ref{eq:aGGshifted}%
). In the case of a pair in the Cooper channel the left and right edges of
the branch-cuts in $\Psi $ play a similar role, see Eqs.~(\ref{eq:psizero})
and~(\ref{eq:nondiagonal}). A diagram in the thermodynamic potential with
the two edges running on each other produces the anomalous temperature
dependence in the spin susceptibility because of its high sensitivity to the
temperature and magnetic field at the point when they touch, see Fig. \ref%
{fig:twogreensfunctions}. The spectral weight of the edge-modes is much
weaker compared to that of the usual quasiparticles, e.g., phonons. However,
taking derivative with respect to an external parameter, a magnetic field in
the case of the spin susceptibility, makes the spectral weight singular. As
a result, the edge modes become active and are able to generate a
non-analytic $f$ at $\omega=0$.

To get a clue of the origin of the discussed non-analytic terms, the
following analogy may be useful. As it has been mentioned, a product of two
propagators of the edge modes running on each other generates non-analytic
terms in the spin susceptibility. With this in mind, consider an ensemble of
one-dimensional right- and left-moving particles turning into each other due
to a mixing matrix element. In such an auxiliary system, the diagrammatic
expansion with respect to the mixing amplitude would suffer from strong
''infrared'' divergences due to the degeneracy of the right- and left-moving
particles at $q=0$. In particular, in the second order in mixing the
thermodynamic potential has a term $\sim V_{mix}^{2}T\sum \int {d}%
q(\omega_{n}^{2}+q^{2})^{-1}$. The avoiding of the level crossing leads to
the restructuring of the ground state of this system with the gap opening at 
$q=0$. Coming back to the non-analytic term in spin susceptibility in 2D at
a finite temperature, we can see that the magnetic field $H$ (multiplied by
the interaction amplitude) acts as the mixing matrix element in the
subsystem of the edge modes. The considered diagrams are not divergent,
because the spectral weight of the edge-modes is much weaker compared to the
usual quasiparticles, e.g., phonons. Therefore, there are no enough
resources for restructuring of the ground state. In addition, being
composite objects, the edge modes are smeared by the temperature through the
Fermi-Dirac function $n_{F}$ of the two fermions from which they are made
of. Nevertheless, they still are able to produce the anomalous terms in the
thermodynamic potential which appear as a trace of the mentioned above
divergences in the expansion with respect to mixing amplitude in the
auxiliary system. The analogy with auxiliary system also helps to understand
the dependence on the order of taking the limits $H\rightarrow0,T\rightarrow
0$. This dependence is similar to the competition between the value of the
mixing matrix element and the energy of the particles in the auxiliary
system, provided that the energy is substituted by the temperature in the
original system of interacting electrons.

In Apppendices we show how the non-analytic (edge) parts in the product of
two pairs of Green's functions generate anomalous dependencies in the
thermodynamic potential. With this experience let us now discuss possible
generalizations of this mechanism. In diagrams defining the spin
susceptibility there are two differentiations with respect to the magnetic
field. The calculation of the term with two pairs indicates that for
nucleating the non-analyticity in the thermodynamic potential the
quasiparticles involved in the differentiation have to be constrained by the
conservation of momentum and energy. For the product of two pairs these
constraints are imposed automatically. Since for the more complicated
diagrams this may not be the case, in identifying other non-analytic terms
it is important to preserve this feature. Consequently, in the thermodynamic
potential only those diagrams are essential that can be arranged as a closed
ladder loop, and where only the Green's function within the sections of the
ladder are differentiated with respect to the magnetic field (by section we
understand a pair of Green's function describing a propagation of a pair of
quasiparticles between the scattering events).

Organizing a diagram in the form of a ladder can be performed as follows.
(In the discussion below the closed ladder loop should be not confused with
the fermionic loops). If the Green functions marked by the differentiation
belong to two different fermionic loops and there exists at least one
another pair such that removing this group of Green's functions
(articulation quadruplet in the terminology of Ref.~%
\onlinecite{Dominicis1964}) will split the diagram into two disconnected
pieces, then these Green's functions nucleate the ladder. Obviously, the
Green's functions that split the diagram are constrained by the conservation
of energy and momentum. When such splitting requires more than one
supporting pair, the non-analytic parts, which are a goal of our studies,
are smeared by the additional integrations and this diagram cannot generate
the linear in temperature term in spin susceptibility. When the two
differentiated Green's functions belong to the same fermionic loop, one
first chooses another fermionic loop and repeats the procedure. Namely, it
is needed to find another supporting pair of Green's functions within the
second loop which allows splitting the diagram into disconnected pieces.
This completes the procedure of arranging the differentiated Green's
functions as a part of a ladder.

Unlike the one-dimensional case, the renormalizations in the Cooper and $%
2p_{F}$ channels are not mutually cross-coupled by the logarithmic
corrections in two dimensions. This makes the Cooper channel the only one in
which terms logarithmically divergent in temperature are generated. Besides
that, in the treatment of the Cooper ladder we can consider its crossbars
while ignoring their non-analytic parts originating from the $2p_{F}$
channel. We rely here on that (i) the non-analytic parts in the $2p_{F}$%
-scattering are small unless they are differentiated (notice that all the
differentiated Green's functions have been attributed to the Cooper ladder
here), and more importantly that (ii) the non-analyticity generated in
another channel is smeared out. That's why for the calculation of the
anomalous terms generated in the Cooper channel it is enough to consider a
ladder assuming that the crossbars in the Cooper ladder are analytic.\cite%
{twodiff} On these grounds, for the analysis of the non-analytic temperature
dependences we can use as a starting point the conventional theory of the
Fermi-liquid systems.

\section{Physical consequences of the renormalizations in the Cooper channel.%
}

Cooper's logarithms lead to renormalizations of the linear in $T$ term in
the spin-susceptibility generated in the Cooper and $2p_{F}$ channels (see
Appendix~\ref{sec:Cooperchannel} for details): 
\begin{equation}
\delta \chi =\nu \frac{T}{\epsilon _{F}}|\Gamma ^{C}(\pi )_{T}|^{2}\,,
\label{eq:ttildeone}
\end{equation}%
where $\nu =m/2\pi $ is the density of states per one species ($m$ is the
effective mass which includes Fermi liquid renormalizations), and we omit
the factors $(g\mu _{B}/2)$ everywhere. The renormalized backward scattering
amplitude $\Gamma ^{C}(\pi )$ is%
\begin{equation}
\Gamma ^{C}(\pi )_{T}=\sum_{n}(-1)^{n}\gamma _{n}(T),\qquad {\gamma }_{n}({T}%
)=\frac{\Gamma _{n}^{C}}{1+\Gamma _{n}^{C}\ln \Lambda /{T}}\,.
\end{equation}%
Here $\gamma _{n}(T)$ are the renormalized Fourier harmonics amplitudes in
the Cooper channel; $\Gamma _{n}^{C}$ are bare amplitudes defined at a large
energy scale $\sim \epsilon _{F}$. This result is in full correspondence
with the renormalization group equation for the angle-dependent amplitude $%
\Gamma ^{C}(\theta )$ in the Cooper-channel. The fact that the amplitudes $%
\gamma _{n}(T)$ undergo strong renormalizations with temperature leads to
important physical consequences which we now discuss.

When the temperature is reduced, the amplitudes $\gamma _{n}(T)$ which have
a positive value are renormalized to zero, while those that initially are
negative grow in magnitude. The initial (bare) value of $\Gamma^{C}(\pi)$ is
most probably dominated by the zero-harmonic amplitude. It is naturally to
expect that this harmonic is repulsive, and therefore when the temperature
is lowered it dies out. In the intermediate region of temperatures when
negative amplitudes are still small, i.e., when $1/|\Gamma _{n}^{C}|>\ln
\Lambda /T>1/\Gamma _{0}^{C}$, the temperature dependence of the spin
susceptibility is 
\begin{equation}
\delta \chi \approx (\nu /\epsilon_{F})\;\frac{T}{(\ln\Lambda /T)^{2}}\,.
\label{eq:logsquare}
\end{equation}%
This result has been also obtained in Ref.~[\onlinecite{georg}] using a
technique developed recently for the ballistic systems.\cite{Efetov2006}

As the temperature is lowered, the negative amplitudes start to grow. There
are general reasons why the negative amplitudes are always present for some
harmonics.\cite{Kohn1965} The largest negative amplitude, $\gamma
_{n}=\gamma ^{C}$, is most important as this amplitude grows most rapidly
approaching the Cooper instability. When the amplitude $|\gamma^{C}(T)|$
becomes of order of unity, it acquires an essential temperature dependence.
At small enough temperatures, when the renormalized to zero positive
harmonics are suppressed, see Eq.~(\ref{eq:logsquare}), the temperature
dependence of the growing negative amplitude $\gamma ^{C}(T)=\Gamma
^{C}/(1+\Gamma ^{C}\ln \Lambda /{T)}$ prevails and can determine the
temperature behavior of the spin susceptibility. This may cause a
non-monotonic behavior in $\delta\chi(T)$. For the purpose of illustration,
we will discuss the temperature dependence in the spin susceptibility
leaving only the most rapidly growing amplitude $\gamma ^{C}(T)$. In this
case the temperature dependence of $\delta \chi (T)\varpropto T\gamma
^{C}(T)^{2}$ is determined by two competing terms:%
\begin{eqnarray}
d\delta \chi /dT &=&(\nu /\epsilon _{F})[\gamma ^{C}(T)^{2}+2T\gamma
^{C}(T)(d\gamma ^{C}(T)/dT)]  \notag \\
&=&(\nu /\epsilon _{F})(\gamma ^{C}(T)^{2}-2|\gamma ^{C}(T)|^{3}).
\label{eq:derivativeT}
\end{eqnarray}%
This expression changes its sign when $-\gamma ^{C}(T)>1/2$.

Let us now consider the present experimental situation.~\cite{Prus2003} The
data indicate a noticeable temperature dependence of the spin susceptibility
at $T\gtrsim2K$, which is too strong to be attributed to the conventional
Fermi-liquid corrections as a possible explanation. On the other hand, the
possibility that this temperature increase is due to the presence of a large
portion of free localized magnetic moments is probably not very realistic;
at metallic densities that are $3\div4$ times higher than the critical
density $n_{c}$, the localized moments should be coupled by the RKKY
interaction. Here we look at the observed temperature dependence from the
point of view of the non-analytic corrections discussed in the paper.

The observed trend of the temperature dependence corresponds to decrease of
the spin susceptibility with temperature. Equation~(\ref{eq:logsquare})
predicts, however, an opposite trend. The explanation of the experiment may
be attributed to the non-monotonic behavior of the spin susceptibility due
to temperature renormalizations of the negative amplitudes $\gamma_{n}$, as
has been discussed above. In Ref.~[\onlinecite{Prus2003}] the data are
presented for $n/\chi (T)$, where $n$ is the density of the 2D electron gas.
For silicon MOSFET (there are two valleys) the density $n=p_{F}^{2}/\pi$.
For the degenerate Fermi gas with two valleys the unrenormalized spin
susceptibility $\chi _{0}=4\nu $ and the ratio $n/\chi _{0}=p_{F}^{2}/2m$.
In the presence of two valleys the expression for $\delta\chi $ as given by
Eq.~(\ref{eq:ttildeone}) should be multiplied by the factor $4$, i.e., $%
\delta \chi =4\nu (T/\epsilon _{F})\gamma ^{C}(T)^{2}$. Since the
temperature corrections are relatively small, one can expand $n/\chi$: 
\begin{equation}
\delta (n/\chi )\approx \frac{n}{\chi _{0}^{2}}(-\delta \chi )=-T\gamma
^{C}(T)^{2}\,.  \label{eq:deltanchi}
\end{equation}%
The main feature of this relation is that the temperature dependence of $%
\delta(n/\chi)$ is determined by the amplitude $\gamma ^{C}(T)$ only. Notice
that the modification of the spin susceptibility by the Stoner factor drops
out from $\delta(n/\chi)$ because of the renormalization of the two external
vertices in $\delta\chi$, which are indicated by triangles in Fig.~\ref%
{fig:4Greenfunc}, see Eq.~(\ref{eq:renormalizedchi}).

The data\cite{Prus2003} exhibit a noticeable increase of $\delta (n/\chi )$
starting from $T\gtrsim2K$ and correspond to $|\gamma ^{C}(T)|\sim 1$ in
Eq.~(\ref{eq:derivativeT}). Such a value of the amplitude $\gamma ^{C}(T)$
in the Cooper channel implies that the experiment has been performed at
temperatures comparable with the temperature of the superconducting
instability. Note, however, that in the disordered system studied in Ref.~[%
\onlinecite{Prus2003}] the superconducting instability cannot fully develop
for a non-zero harmonic ${\gamma }_{n}({T})=1/[(1/\Gamma _{n}^{C})^{-1}+\ln
(\Lambda /\max {T,\tau }_{elastic}^{-1}){]}$, because the instability is
blocked by disorder when ${\tau }_{elastic}^{-1}\gtrsim T_{c}$; here $T_{c}$
is the temperature of the Cooper instability in the dominant harmonic.

The experimental curves well in the metallic region correspond to $%
\epsilon_{F}\sim30\div40K$. We evaluate the bare value of the amplitude $%
\gamma^{C}$ as $\Gamma ^{C}\sim -0.25\div -0.3$. Then, $\gamma ^{C}(T)$
reaches the value $\gamma ^{C}(T)=0.5,$ where $\delta \chi $ changes its
sign, at temperatures $\sim 10K$. Would it not be blocked by disorder, the
superconducting instability may occur at $T_{c}\sim 1K.$ The measurements
are performed up to $T\lesssim 4K$. In the suggested explanation, the
initial decrease with temperature of the spin susceptibility should be
succeeded by an increase at larger temperatures, $T\gtrsim 10K$. The
temperature range of the existing measurement is too small, however, to make
this conclusions definite.

Two comments may be in place here. (i) The analysis of $\delta \chi (T)$
basing on Eq.~(\ref{eq:derivativeT}) is only qualitative when $|\gamma
^{C}(T)|\sim 1$. In fact, the situation is more complicated. The derivative $%
\partial \chi /\partial T$ is a series in $\gamma _{l}(T)$ which does not
reduce to Eq.~(\ref{eq:derivativeT}): 
\begin{align}
& d\delta \chi /d{T}=(\nu /\epsilon _{F})\sum \Big[(-1)^{l^{\prime
}+l^{\prime \prime }}\gamma _{l^{\prime }}({T})\gamma _{l^{\prime \prime }}({%
T})  \notag \\
& +c_{3}^{l^{\prime }l^{\prime \prime }l^{\prime \prime \prime }}\gamma
_{l^{\prime }}({T})\gamma _{l^{\prime \prime }}({T})\gamma _{l^{\prime
\prime \prime }}({T})+\cdots \Big]\,.  \label{eq:notrg1}
\end{align}%
The details are in the end of Appendix~\ref{sec:Cooperchannel}. To get a
final conclusion concerning a non-monotonic behavior, one has to calculate
for $|\gamma ^{C}(T)|\sim 1$ the whole series in Eq.~(\ref{eq:notrg1}),
which is hardly possible. (ii) Initial values of the non-zero harmonics in
the Cooper channel are highly non universal. In addition, the destructive
influence of the disorder can stop the renormalization of the amplitude of
the dominant harmonic before it becomes noticeable. Taken together, these
facts mean that the experimental situation may strongly vary from sample to
sample.

In a separate publication \cite{PNAS2006}, we present an alternative explanation of the experimental situation. We consider repulsive scattering amplitudes only, $\Gamma_n^C>0$, that is, perhaps, more appropriate for a system studied in Ref.~[\onlinecite{Prus2003}]. Since the repulsive scattering amplitudes scale to zero, a more delicate analysis is necessary there. One has to account for all three channels, i.e. to include the rescattering of particle-hole pairs in the zero-sound channel in addition to the Cooper and $2p_F$-scattering channels. In this way it is possible to explain the observed temperature dependence of the spin susceptibility, both in sign and temperature. (Note that the non-monotonic behavior may also occur here. The temperature dependence changes its sign when the logarithmic suppression of the repulsive amplitudes become ineffective, i.e., when $1/\ln(\epsilon_F/T)\gtrsim\gamma^C$.)

\section{Concluding remarks.}
\label{sec:remarks}

A decade ago, the authors of Ref.~[\onlinecite{Belitz1996b}] made an
important conjecture, based on the power counting, that thermodynamic
quantities in the 2D electron gas contain linear in $T$ dependences. They
argued that similarly to one dimension, where the spin susceptibility
depends logarithmically on the temperature,\cite{Larkin1973a} the anomalous
temperature terms should also exist in higher dimensions, albeit in a weaker
form.

We believe that presented calculation of the spin susceptibility in 2D does
not support this line of reasoning. Indeed, the temperature dependence of
the spin susceptibility in one dimension (1D) originates from the
logarithmic in temperature dependence of the Stoner enhancement calculated
in the parquet renormalization scheme.\cite{Larkin1973a,Affleck1994} In
contrast to the Fermi liquid description (which is analyzed in terms of a
propagation of two quasi-particles) and the parquet approximation (which is
analyzed with the Sudakov cross-section of two Green's functions), the\
anomalous terms in the thermodynamic potential in 2D develops only when two
(or more) edge parts of the two-particle correlation functions act in the
combined way. In particular, in 2D, unlike 1D, the temperature dependence of
the spin susceptibility does not reduce to a mere renormalization of the
Stoner factor. In our opinion, the physics of the temperature corrections in
2D is rather unique and is not a mere continuation of 1D logarithms.

It has been shown in Appendices~\ref{sec:twoGreen}~and~\ref{sec:hnm} that
both a $2p_{F}$-pair and a two-particle section in the Cooper channel have
non-analytic parts (edge parts). In two dimensions the square root
singularities of these edge parts are relatively weak. However, taking
derivatives with respect to an external parameter, a magnetic field in our
case, makes them more singular, i.e., activates them. Because of high
sensitivity to the temperature and magnetic field, diagrams containing edge
parts running on each other produce the anomalous terms in the spin
susceptibility. When a diagram for the thermodynamic potential consists only
of two pairs of quasiparticles, their edge parts are effective because the
quasiparticles are constrained by the conservation of the momentum and
energy. In this case the edge parts of the two pairs act together in a
combined way. For more complicated diagrams the non-analytic pieces may be
not constrained. Then they are smeared out by independent integrations and
become ineffective. This leads us to the conclusion that for the anomalous
temperature dependences the essential diagrams in the thermodynamic
potential are ladder diagrams only. For the spin susceptibility the set of
essential diagrams in the $2p_{F}$ and Cooper channels is organized in a way
that the anomalous terms can be considered within the Cooper channel only. (This rearranging is necessary to avoid the double-counting of the two-section terms which, as has been found here, completely overlap in the $2p_F$  and Cooper channels.) 
We show next that one has to keep arbitrary number of sections in the Cooper
channel ladder to collect all non-analytic terms in the spin susceptibility,
i.e., not only a product of the two pairs. (This is because the singular
part of a two-particle section in the Cooper channel depends on the ratio $%
\omega/qv_{F}$.) The truncation of the series in powers of the renormalized
amplitudes is possible only when these amplitudes are small.

In Refs.~[\onlinecite{Galitski2005,GlazmanB2005}] the possibility of strong
renormalizations in the Cooper channel was neglected and the amplitude $%
\Gamma (\pi )$ was analyzed from the point of view of the Fermi-liquid
renormalizations. These renormalizations are not related to the infrared
logarithms, but they may be large near Pomeranchuk's instability in one of
the channels, e.g., near the Stoner instability in the case of the magnetism.%
\cite{Galitski2005} Since the amplitude $\Gamma (\pi )$ involves all angular
harmonics, an instability in any of them can be important. In this paper we
assumed that the discussed Fermi liquid system is not in the immediate
proximity of any instability. In this respect our analysis emphasizing only
the infrared renormalizations in the Cooper channel may be incomplete.

Finally, let us point out that when the Fourier harmonics amplitudes in the Cooper channel, $\Gamma_n^C$, are repulsive and scale to zero at low temperatures, the contributions to the spin susceptibility from the Cooper channel become unimportant. This situation is analyzed in a separate publication \cite{PNAS2006}, which is complementary to the present one. For vanishing Cooper amplitudes, the anomalous temperature terms are dominated by the non analytic contributions from the particle-hole pairs with small momentum transfer (zero-sound channel). Extending our analysis to particle-hole rescattering, we show that, exactly like in the Cooper channel, the two-section term in the zero-sound channel is dominated by the backward scattering. To avoid the double counting which now involves all three channels, the two-section term has been fully attributed to the Cooper channel, where it gets killed off by the logarithmic renormalizations of the repulsive amplitudes. We thus conclude that for the repulsive interaction the anomalous temperature corrections are determined by the ladder diagrams in the zero-sound channel with three and more rescattering sections.\cite{threesections}

\appendix

\section{Analytic properties of the product of two Green's function with $%
q\approx2p_F$}

\label{sec:twoGreen}

Consider a product of two Green's functions summed over the momentum and the
energy as a function of the transferred energy $i\omega _{n}$ and momentum $%
\mathbf{q}$ 
\begin{equation}
\lbrack GG]_{i\omega _{n},\mathbf{q}}=\int \frac{d^{2}p}{(2\pi )^{2}}\frac{%
n_{F}(\epsilon _{\mathbf{p}+\mathbf{q}}-\mu )-n_{F}(\epsilon _{\mathbf{p}%
}-\mu )}{\epsilon _{\mathbf{p}+\mathbf{q}}-\epsilon _{\mathbf{p}}-i\omega
_{n}}\,,  \label{eq:GG1}
\end{equation}%
where $n_{F}$ is the Fermi-Dirac distribution function $n_{F}(\epsilon
)=[\exp (\beta \epsilon )+1]^{-1}$. We are interested in the analytic
structure of Eq.~(\ref{eq:GG1}) in the complex-$\omega $ plane when $q\equiv
|\mathbf{q}|\approx 2p_{F}$. To get an idea about the analytic properties of
a $2p_{F}$-pair of Green's functions, let us start from the case when $%
\epsilon _{\mathrm{p}}=(p_{\parallel }^{2}+p_{\perp }^{2})/2m$. We first
shift the momentum $\mathbf{p}+\mathbf{q}$ to $\mathbf{p}$ in the first term
in the numerator and write the $d^{2}p$-integration as $d^{2}p=dp_{\parallel
}dp_{\perp }$, where $p_{\parallel }$ is along the direction of $\mathbf{q}$%
. Then, the variable $p_{\perp }$ drops from the denominator and enters only
via $n_{F}(\epsilon _{p}-\mu )$. This parametrization allows us to introduce
the ''spectral density'' $\rho (p_{\parallel })$ which determines the
dependence on the chemical potential and the temperature 
\begin{equation}
\rho (p_{\parallel })=\int_{-\infty }^{+\infty }\frac{dp_{\perp }}{2}%
n_{F}(\epsilon _{\mathbf{p}}-\mu )\,.  \label{eq:spectraldensity}
\end{equation}%
After that get an expression for the $2p_{F}$-pair in the form of the
Lehmann-like representation in the complex-$\omega $ plane 
\begin{equation}
\lbrack GG]_{\omega ,\mathbf{q}}=\frac{m^{1/2}}{4\pi v_{F}}\left( \Pi
_{l}(q,\omega )-\Pi _{r}(q,\omega )\right) \,,
\end{equation}%
where the functions $\Pi _{l}(q,\omega )$ and $\Pi _{r}(q,\omega )$ are
defined as 
\begin{equation}
\frac{m^{1/2}}{4\pi v_{F}}\Pi _{l,r}(q,\omega )=\frac{m}{2\pi q}\int \frac{%
dp_{\parallel }}{\pi }\frac{\rho (p_{\parallel })}{p_{\parallel
}-(m/q)(\omega \pm q^{2}/2m)}\,.  \label{eq:Lehmann}
\end{equation}%
Here $\pm $ in the denominator corresponds to $\Pi _{l}$ and $\Pi _{r}$,
respectively. From now on, in this Appendix we consider the temperature to
be zero. [The non-zero temperature can be restored with the help of
equation~(\ref{eq:relationforPi}) as it is explained below.] At $T=0$ the
integral in Eq.~(\ref{eq:spectraldensity}) is easily evaluated: $\rho
(p_{\parallel })_{T=0}=\sqrt{p_{F}^{2}-p_{\parallel }^{2}}\theta
(p_{F}-p_{\parallel })$. In the complex-$\omega $, plane $\Pi _{l,r}(\omega
) $ considered as a function of the variable $u_{\pm }=(m/qp_{F})(\omega \pm 
{q}^{2}/2m)$, have the following analytic form 
\begin{align}
& \frac{m^{1/2}}{4\pi v_{F}}\Pi _{l,r}=\frac{mp_{F}}{2\pi q}F(u_{\pm }); 
\notag \\
& \qquad F(u)=\int_{-1}^{1}\frac{dx}{\pi }\frac{\sqrt{1-x^{2}}}{x-u}=-u+%
\sqrt{u-1}\sqrt{u+1}\,.  \label{eq:Ffunction}
\end{align}%
Here the square roots should be understood as having a branch cut directed
to the right on the real axis. Correspondingly, $F(u)$ has a branch cut
between $-1$ and $1$ in the complex-$u$ plane. On the upper side of the
branch cut the imaginary part of $F(u)$ is positive; the real part
originating in the square root term is positive to the right of the branch
cut and is negative to the left of it.

\begin{figure}[tbp]
\centerline{\includegraphics[width=0.35\textwidth]{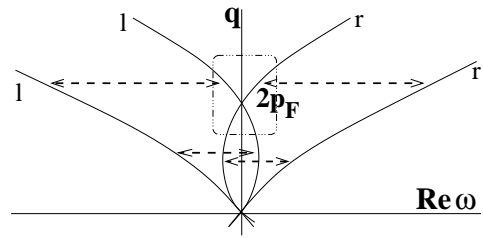}}
\caption{Illustration of the analytic structure of the $2p_F$ pair of
Green's functions. The horizontal axis is $\Re\protect\omega$ and the
vertical axis is $q$. The parabolas show the position of the edges of the
branch cuts of $\Pi_{l,r}$ in the $q,\mathrm{Re}\protect\omega$-plane. The
dashed lines show the branch cuts of $\Pi_{r,l}$ for two typical values of $%
q $: the $\Pi_r$-part has branch cuts between the two right-looking
parabolas, and $\Pi_l$ has branch cuts between the two left-looking
parabolas. The rectangle marks the region where the edges of the two branch
cuts of two $2p_F$ pairs are close to each other. The meeting of two edges
leads to non-analytic behavior of the thermodynamic potential.}
\label{fig:twogreensfunctions}
\end{figure}

Only the second term in $F(u)$ containing the square roots can be
responsible for non-analytic behavior in the thermodynamic potential. The
non-analytic temperature dependence in the product of four Green's functions
originates from region marked by dashed-line rectangle in Fig.~\ref%
{fig:twogreensfunctions} where the edges of the branch cuts of $\Pi_r$ and $%
\Pi_l$ are close to each other. This makes the contribution from this region
very sensitive to the magnetic field and temperature smearing.

For the calculation of the non-analytic term in the spin susceptibility it
is enough to use an appropriate approximation for $2p_{F}$-pair in the
region marked by a dashed-line rectangle in the Fig.~\ref%
{fig:twogreensfunctions}. This involves the $u=-1$ edge of branch cut in $%
\Pi _{r}$ and the $u=1$ edge of the branch cut in $\Pi _{l}$. In the
following we omit the smooth terms in $F(u)$. It can be checked by an
explicit calculation that the first term in $F(u)$ does not to contribute to
the anomalous temperature behavior.

The spectrum of electrons is not considered to be quadratic, like in Eqs.~(%
\ref{eq:Lehmann})~and~(\ref{eq:Ffunction}), but we still assume the
spherical symmetry. The generalization for a non-spherical spectrum can be
done, if needed. To devise an adequate approximation for $\Pi _{lr}$, we
consider again the product of two Green's functions, $\big[G_{+}G_{-}\big]%
_{\delta q,i\omega }$ describing the propagation of two quasiparticles with
the momenta $\mathbf{p}_{+}$ and $\mathbf{p}_{-}$, which are almost opposite
to each other. The transferred momentum $\mathbf{q}=\mathbf{p}_{+}-\mathbf{p}%
_{-}$ is close to $2p_{F}$, and $\delta q=|\mathbf{q}|-2p_{F}$ describes a
small deviation from $2p_{F}$. We choose the following parametrization where 
$\mathbf{p}_{F}$ is a vector directed along $\mathbf{q}$: 
\begin{align}
\mathbf{q}\;\;=& \;\;2\mathbf{p}_{F}+\delta \mathbf{q},  \notag \\
\mathbf{p_{-}}=& -\mathbf{p}_{F}+\delta \mathbf{p}\,.
\end{align}%
To reproduce the correct threshold behavior we keep minimal terms in the
expansion in $p_{\parallel }$ and $p_{\perp }$. We linearize the energy
spectrum around the points $\pm \mathbf{p}_{F}$ in the direction
perpendicular to the Fermi surface, but preserve the quadratic momentum
dependence in the direction parallel to the Fermi surface to account for the
Fermi surface curvature 
\begin{align}
\epsilon _{\mathbf{p}_{-}}-\mu _{0}=& -v_{F}\delta p_{\parallel }+p_{\perp
}^{2}/2m  \notag \\
\epsilon _{\mathbf{p}_{+}}-\mu _{0}=& \;\;v_{F}(\delta p_{\parallel }+\delta
q)+p_{\perp }^{2}/2m\,.  \label{eq:linearization}
\end{align}%
We obtain the following approximation for $\Pi _{r,l}(\omega )$ at zero
temperature 
\begin{align}
& \frac{m^{1/2}}{4\pi {v}_{F}}\Pi _{l,r}(q,\omega )  \notag \\
& \;\;=\frac{1}{2\pi }\int \frac{dp_{\parallel }dp_{\perp }}{2\pi }\;\frac{%
n_{F}\Big(\pm {v}_{F}(\delta p_{\parallel }+\delta q)+p_{\perp
}^{2}/2m-\delta \mu \Big)}{v_{F}(2\delta p_{\parallel }+\delta q)-\omega }\,,
\label{eq:GGlinearized}
\end{align}%
where $\pm $ corresponds to $\Pi _{l}$ and to $\Pi _{r}$, respectively.

Having in mind the calculation of the spin susceptibility from the
correction to the thermodynamic potential we introduce a shift of the
chemical potential $\mu =\mu _{0}+\delta \mu $ in Eq.~(\ref{eq:GGlinearized}%
). Introducing $x=2v_{F}(\delta p_{\parallel }+\delta {q})\pm 2\delta \mu $
in $\Pi _{r}$ and $\Pi _{l}$, respectively, we obtain after integrating over 
$p_{\perp }$ the following expressions for $\Pi _{l,r}$ 
\begin{align}
& \Pi _{l}(\delta {q},\omega )_{T=0}=\int \frac{dx}{\pi }\frac{\sqrt{-x}%
\theta (-x)}{x-(\omega +v_{F}\delta q-2\delta \mu )}  \notag \\
& \Pi _{r}(\delta {q},\omega )_{T=0}=\int \frac{dx}{\pi }\frac{\sqrt{x}%
\theta (x)}{x-(\omega -v_{F}\delta q+2\delta \mu )}\,.  \label{eq:aGGshifted}
\end{align}%
These expressions correspond to Eq.~(\ref{eq:Lehmann}) within the region
marked by dashed-line rectangle in Fig.~\ref{fig:twogreensfunctions}. Here
the functions $\Pi _{l,r}$ are defined with positive square roots in the
numerator. In this Lehmann-like representation, $\Pi _{l}$ describes the
square root function with the branch cut directed to the left on the real
axes of $\omega $-plane, while $\Pi _{r}$ describes the function with the
branch cut directed to the right. [Indices $l$ and $r$ correspond also to
the direction of ''propagation'' in the combinations $\omega \pm {v}_{F}q$.]
Both $\Pi _{l}$ and $\Pi _{r}$ have a positive imaginary part on the upper
side of their branch cuts in the complex $\omega $-plane.

Let us conclude this Appendix by a comment on Eq.~(\ref{eq:aGGshifted}).
This equation gives a representation of the edge parts of $2p_{F}$-pair of
Green's functions as two pieces, $\Pi _{r}$ and $\Pi _{l}$, in the region
inside the dashed-line rectangle in Fig.~\ref{fig:twogreensfunctions}.
Together these edge parts act as a sort of an ''edge-mode''; we mean the
combination $\omega \pm {v}_{F}(q-2p_{F})$ in the denominators of Eq.~(\ref%
{eq:aGGshifted}). The square root spectral weight of these edge-modes is
much weaker compared to the usual quasiparticles, e.g., phonons. However,
taking derivative with respect to an external parameter, a magnetic field in
our case, makes the spectral weight singular. As a result, these edge modes
become active. In Appendices~\ref{sec:2pf}~and~\ref{sec:Cooperchannel} we
show that, because of the unusual sensitivity to the temperature and
magnetic field, a diagram in the thermodynamic potential with two
propagators of the edge modes running on each other produces the anomalous
temperature dependence in the spin susceptibility of the interacting
electron gas.

\section{Anomalous terms generated by two $2p_{F}$-pairs.}

\label{sec:2pf}

The term in the thermodynamic potential with four Green's functions which
produces the anomalous term in the spin susceptibility is 
\begin{equation}
\delta \Omega _{2}(T,h)=-(1/2)g_{4}^{\uparrow \downarrow }(T;h,-h)|\Gamma
(\pi )|^{2}\,,  \label{eq:anomalous}
\end{equation}%
where $g_{4}(\delta \mu _{1},\delta \mu _{2};T)$ signifies the product of
the two pairs of Green's functions with the chemical potential shifted in
two pairs by $\delta \mu _{1}$ and $\delta \mu _{2}$, correspondingly;
arrows indicate spin and we consider a Zeeman energy $h=(g\mu _{B}/2)H$ as
the source of a shift of the chemical potentials. The factor $1/2$ can be
obtained from the comparison with the second order term in the perturbation
theory for the thermodynamic potential. [There are two such contributions,
with $\uparrow \downarrow $ and $\downarrow \uparrow $ arrangements of spins
in the two pairs of Green's functions, which has been accounted for by the
factor $1/2$ in Eq.~(\ref{eq:anomalous})]. The terms with the parallel spin
projections in the two pairs of Green's functions, i.e., $g_{4}^{\uparrow
\uparrow }(T;h,h)$ and $g_{4}^{\downarrow \downarrow }(T;h,h)$, do not
contribute to the anomalous term.

Having in mind the correction to the thermodynamic potential $\delta \Omega
_{2}$ given by Eq.~(\ref{eq:anomalous}), we now perform the sum over
Matsubara frequency $i\omega _{n}$. The standard manipulation leads to 
\begin{align}
& g_{4}(T;\delta \mu _{1},\delta \mu _{2})=\frac{mp_{F}}{(4\pi v_{F})^{2}}%
\int \frac{d\delta q}{\pi }\int \frac{d\omega }{2\pi }\coth \frac{\beta
\omega }{2}\,  \notag \\
& \Big\{\Im \big[\Pi _{l}(\delta \mu _{1})-\Pi _{r}(\delta \mu _{1})\big]%
_{T}\;\Re \big[\Pi _{l}(\delta \mu _{2})-\Pi _{r}(\delta \mu _{2})\big]_{T} 
\notag \\
& \;+(\delta \mu _{1}\leftrightarrow \delta \mu _{2})\Big\}\,.
\label{eq:derivation1}
\end{align}%
The signs of different terms in this expression are completely determined by
the analytical properties of functions $\Pi _{r,l}$. Only the cross-product
terms, i.e., $\Pi _{l}\Pi _{r}$, behave sharply in the vicinity of $\omega
=0 $, and therefore are responsible for non-analytic terms in the
thermodynamic potential. It can be checked directly that in a product of,
let us say, $\Pi _{r}$ with a regular function its square root singularity
disappears after the $q$-integration. For similar reasons, the products $\Pi
_{l}\Pi _{l}$ and $\Pi _{r}\Pi _{r}$ yield regular functions in frequency.
On the contrary, in the product $\Pi _{l}\Pi _{r}$ the edges of the two
branch cuts run on each other, preventing the smearing of the singularity.

The temperature dependence in the functions $\Pi _{l,r}$ can be restored
with the help of the relation 
\begin{equation}
\Pi (\omega )_{T}=\int ds\left( -\frac{\partial n_{F}(s)}{\partial s}\right)
\Pi (\omega -s)_{T=0}  \label{eq:relationforPi}
\end{equation}%
which follows from the identity 
\begin{equation}
n_{F}(\omega )_{T}=\int ds\left( -\frac{\partial n_{F}(s)}{\partial s}%
\right) n_{F}(\omega -s)_{T=0}\,.  \label{eq:identity}
\end{equation}%
Performing the $q$-integration in Eq.~(\ref{eq:derivation1}) one gets 
\begin{align}
& g_{4}(\delta \mu _{1},\delta \mu _{2})_{T}=-\frac{\nu ^{3}}{8\epsilon _{F}}%
\int d\omega ds_{1}ds_{2}\frac{\partial n_{F}(s_{1})}{\partial s_{1}}\frac{%
\partial n_{F}(s_{2})}{\partial s_{2}}\,\times  \notag \\
& \quad \coth \frac{\beta \omega }{2}\Big[(\omega +\delta \mu _{1}-\delta
\mu _{2}+s_{1}-s_{2})^{2}\mathrm{sign}(\omega +\Delta +s_{1}-s_{2})\Big]\,,
\label{eq:gfunction}
\end{align}%
The obtained expression depends on the relative shift $\Delta =\delta \mu
_{1}-\delta \mu _{2}$ of the chemical potential in the two pairs of Green's
functions. The \emph{signum} function in this expression appears because the
contributions with $\Im \Pi _{r}(\delta \mu _{1})\;\Re \Pi _{l}(\delta \mu
_{2})$ and $\Re \Pi _{r}(\delta \mu _{1})\;\Im \Pi _{l}(\delta \mu _{2})$
have opposite signs. The expression~(\ref{eq:gfunction}) contains the needed
non-analytic frequency dependence through the factor $\mathrm{sign}(\omega
+\delta \mu _{1}-\delta \mu _{2})$. This factor compensates the
odd-in-frequency behavior of $\coth $ allowing for the cubic $\Delta ^{2}T$
term in the expansion of $g_{4}(\Delta ,T)$ which produces the linear in $T$
term in the spin susceptibility. Notice that the expression obtained in Eq.~(%
\ref{eq:gfunction}) explains how the dependence on the order of taking the
limits $H\rightarrow 0,T\rightarrow 0$ appears in the thermodynamic
potential. At $T\ll H$ the temperature dependent terms in the thermodynamic
potential vanishes as $T^{3}$. To calculate the spin susceptibility one has
to take the opposite limit, $H\ll T$.

Let us first consider a contribution to the transverse spin susceptibility,
as illustrated in Fig.~\ref{fig:4Greenfunc}. Notice that there are no
''drag'' diagrams for the transverse spin susceptibility, because Green's
functions attached to the external vertices $\sigma ^{\pm }$, carry opposite
spins. The spin susceptibility correction which corresponds to these
diagrams is 
\begin{equation}
\delta \chi =\,2(\chi _{0}/\nu )\,\left( \frac{1}{2}\frac{\partial
^{2}g_{4}(T;\delta \mu _{1},\delta \mu _{2})}{\partial \delta \mu _{1}^{2}} %
\Big\vert_{%
\begin{array}{c}
{\scriptscriptstyle\delta \mu _{1}=0} \\ 
{\scriptscriptstyle\delta \mu _{2}=0}%
\end{array}%
}\right) \,|\Gamma (\pi )|^{2}\,.  \label{eq:transversespin}
\end{equation}%
Here the ratio $(\chi _{0}/\nu )$ accounts for the factors arising in the
calculation of the spin susceptibility from the corresponding diagram with $%
\sigma ^{+}$ and $\sigma ^{-}$ at the vertices. Factor $2$ accounts for a
spin trace in the upper bubble in the diagrams presented in Fig.~\ref%
{fig:4Greenfunc}. The derivative in the brackets generates an expression
with six Green's functions corresponding with the correct coefficient.

The $\omega $ integration in Eq.~(\ref{eq:transversespin}) can be cast in
the form 
\begin{align}
& \frac{1}{2}\frac{\partial ^{2}g_{4}(T;\delta \mu _{1},\delta \mu _{2})}{%
\partial \delta \mu _{1}^{2}}\Big\vert_{%
\begin{array}{c}
{\scriptscriptstyle\delta \mu _{1}=0} \\ 
{\scriptscriptstyle\delta \mu _{2}=0}%
\end{array}%
}=  \notag \\
& \qquad -\frac{\nu ^{3}}{8\epsilon _{F}}\int_{-\Lambda }^{\Lambda }d\omega
-\int d\omega \frac{2-\beta \omega \coth \frac{\beta \omega }{2}}{\cosh
\beta \omega -1}  \label{eq:ultravioletone}
\end{align}%
The first term here is proportional to the ultraviolet cutoff. Since it does
not depend on the temperature, it will be not considered here. [In fact, the
traces of this term reveal themselves in the $\Gamma ^{3}$ terms which
cannot be reduced to the renormalization of the amplitude $\Gamma (\pi )$ by
the rescattering in the Cooper channel.] Omitting the factor $(g\mu
_{B}/2)^{2}$, we obtain for the anomalous contribution in the spin
susceptibility 
\begin{equation}
\delta \chi =\nu \,|\nu \Gamma (\pi )|^{2}\,\frac{T}{\epsilon _{F}}\,,
\label{eq:linearTcorrection}
\end{equation}%
which coincides with the result of previous calculations, see Refs.~[%
\onlinecite{Galitski2005,GlazmanB2005}].

Let us clarify the role of the amplitude $\Gamma (0)$ with the scattering
angle $\theta -\theta ^{\prime }\approx 0$ in the linear in $T$ term in the
spin susceptibility. In the diagrams with two pairs of Green's functions,
which can also be looked at as two pairs with the momentum transfer close to 
$2p_{F},$ the scattering amplitude $\Gamma (0)$ can be mixed with $\Gamma
(\pi )$ or used twice alone. With one exception, these are one-loop diagrams
which cannot contribute to the spin susceptibility (all spin projections are
the same and they don't have part with a difference in the chemical
potential). The only exception is the diagram with two loops. In each loop
the two Green's functions have close momenta, but their directions are
opposite in two loops; the spin projections are also opposite in the two
loops. When rearranged as two pairs with the momentum transfer close to $%
2p_{F}$, they have needed shifts of the chemical potentials, but still $%
\delta \mu _{1}=\delta \mu _{2}$, and therefore the corresponding $%
g_{4}(\delta \mu _{1},\delta \mu _{2})$ does not contribute to the spin
susceptibility. With this comment we fixed the arrangement of the dashed
lines in Fig.~\ref{fig:4Greenfunc}. We see that the magnetic field
dependence drops out from $g_{4}$ unless there are two loops, each with a $%
2p_{F}$-pair of Green's function, and with the spin projections directed
oppositely in each of the loops.

The performed calculation of the anomalous term in the spin susceptibility
demonstrates that it is generated by the pole parts of Green's functions
close the Fermi surface. It follows from this fact that the Fermi liquid
renormalizations of the $\delta \chi $ originating from the two external
vertices (triangles in the Fig.~\ref{fig:4Greenfunc}) are equal to the
square of the Fermi liquid parameter $1/(1+G_{0})$ responsible for the
renormalization of the Pauli spin susceptibility. Including these
renormalizations the temperature correction takes the form 
\begin{equation}
\delta \chi =\frac{\nu }{(1+G_{0})^{2}}|\nu \Gamma (\pi )|^{2}\,\frac{T}{%
\epsilon _{F}}\,.  \label{eq:renormalizedchi}
\end{equation}

The longitudinal spin susceptibility can also be obtained in the same way by
taking the second derivative with respect to the magnetic field, $%
\chi=-\partial^2\Omega_2/\partial h^2$. Notice that in the case of the
transverse spin susceptibility the coefficient $2(\chi_0/\nu)$ originates
from two independent spin traces in Fig.~\ref{fig:4Greenfunc}. In the case
of the longitudinal spin susceptibility there is no spin traces anymore, but
the additional factor 4 in this case originates from the differentiation of $%
g_4(T; h,-h)$ in Eq.~(\ref{eq:anomalous}) with respect to $h$ rather than
with respect to its first argument only as in Eq.~(\ref{eq:transversespin}).
As a result one obtains the same expression as in Eq.~(\ref%
{eq:linearTcorrection}).

\section{Non-analytic terms in the Cooper channel}

\label{sec:Cooperchannel}

In Appendix ~\ref{sec:2pf} it has been assumed from the very beginning that
the transferred momentum $q$ is close to $2p_{F}$. We start the present
Appendix with an alternative treatment of the term $\delta \Omega _{2}$
which now will be recalculated as two sections in the Cooper channel. Here
we will not assume a priory that only backward scattering is important.
Remarkably, we obtain this fact as a result of independent calculation.
Finally, we extend the consideration of the anomalous terms by analysis of
the Cooper ladder in the thermodynamic potential.

\subsection{Calculation of two pairs in the Cooper channel.}

We start with a pair of Green's functions in the Cooper channel 
\begin{align}
& [G(\delta \mu _{1})G(\delta \mu _{2})]_{q,i\omega _{n}}  \notag \\
=& \int \frac{d^{2}p}{(2\pi )^{2}}T\sum_{i\epsilon _{n}}G_{\mathbf{p}+%
\mathbf{q}/2}(i\epsilon _{n}+i\omega _{n};\delta \mu _{1})\;G_{-\mathbf{p}+%
\mathbf{q}/2}(-i\epsilon _{n};\delta \mu _{2})  \notag \\
=& \frac{1}{2}\int \frac{d^{2}p}{(2\pi )^{2}}\frac{\tanh \frac{\beta }{2}%
\big(\xi _{\mathbf{p}+\mathbf{q}/2}-\delta \mu _{1}\big)+\tanh \frac{\beta }{%
2}\big(\xi _{-\mathbf{p}+\mathbf{q}/2}-\delta \mu _{2}\big)}{\xi _{\mathbf{p}%
+\mathbf{q}/2}+\xi _{-\mathbf{p}+\mathbf{q}/2}-\delta \mu _{1}-\delta \mu
_{2}-i\omega _{n}}\,.  \label{eq:Arcady}
\end{align}%
{where $\xi _{\mathbf{p}}=\epsilon _{\mathbf{p}}-\mu $ and $q=|\mathbf{q}|$.}
To account for the effect of magnetic field, we introduced the shift of the
chemical potential $\delta \mu =\pm {h}$ depending on the sign of spin
projection. We now shift the energy variable {$\xi _{\mathbf{p}}$} by $%
(\delta \mu _{1}+\delta \mu _{2})/2$. The obtained expression depends on the
relative shift $\Delta =\delta \mu _{1}-\delta \mu _{2}$ of the chemical
potential in the two Green's functions. Only the pairs with the opposite
spin projections retain the dependence on the magnetic field $h$; the
anomalous temperature terms are not sensitive to the overall shift of the
chemical potential. Finally, after making analytical continuation to the
complex-$\omega $ plane $i\omega _{n}\rightarrow \omega $ we obtain: 
\begin{align}
& [G(\delta \mu _{1})G(\delta \mu _{2})]_{q,\omega }  \notag \\
=& \frac{1}{2}\int \frac{d^{2}p}{(2\pi )^{2}}\frac{\tanh \frac{\beta }{2}%
\Big(\xi +(qv_{F}/2)\cos \theta -\Delta /2\Big)}{2\xi -\omega }+  \notag \\
\phantom{=}& +\Delta \leftrightarrow -\Delta \,,
\end{align}%
{where $\theta $ is the angle between the momenta $\mathbf{p}$ and $\mathbf{q%
}$ in Green's functions in Eq.~(\ref{eq:Arcady}); here we expand $\xi _{%
\mathbf{p}+\mathbf{q}/2}$ using the smallness of }$q$ and introduce $\xi
=\xi _{\mathbf{p}}$; i.e., {$\xi _{\mathbf{p}+\mathbf{q}/2}=\xi
+(qv_{F}/2)\cos \theta $.} The result reproduces the Cooper logarithm
multiplied by the single particle density of states (per spin). We use a
variable $x=2\xi $ to obtain the expression in form of the Lehmann-like
representation: 
\begin{equation}
\lbrack GG]_{q,\theta ,\omega }=\frac{\nu \pi }{4}\Big(\Psi (q,\theta
,\omega +\Delta )_{T}+\Psi (q,\theta ,\omega -\Delta )_{T}\Big)\,,
\end{equation}%
where 
\begin{equation}
\Psi (q,\theta ,\omega \pm \Delta )_{T}=\int \frac{dx}{\pi }\frac{\tanh 
\frac{\beta }{4}\big(x\pm \Delta -qv_{F}\cos \theta \big)}{x-\omega }\,.
\label{eq:Lehman-Psi}
\end{equation}%
The normalization of $\Psi (\theta )$ is such that the imaginary part of $%
\Psi $ approaches $\pm 1$ at large values of $\omega $. At zero temperature
the hyperbolic tangent becomes the \emph{signum} function. The temperature
width of Fermi-Dirac function can be restored from $T=0$ function with the
help of the relation 
\begin{equation}
\Psi (\omega )_{T}=\frac{1}{2}\int ds\frac{\partial \tanh \frac{\beta s}{4}}{%
\partial {s}}\Psi (\omega -s)_{T=0}  \label{eq:fermi-smearing-Psi}
\end{equation}%
that follows from the identity 
\begin{equation}
\tanh \alpha {x}=\frac{1}{2}\int ds\frac{\partial \tanh \alpha {s}}{\partial 
{s}}\mathrm{sign}(x-s)\,.  \label{eq:fermi-smearing}
\end{equation}

The contribution to the thermodynamic potential from two sections (recall
that by section we understand a pair of Green's function describing a
propagation of two quasiparticles between the scattering events) in the
Cooper channel is given by 
\begin{align}
& \Omega _{2}(\Delta )=-\frac{1}{2}\left( \frac{\pi }{2}\right) ^{2}\int 
\frac{qdq}{2\pi }\frac{d\theta _{1}}{2\pi }\frac{d\theta _{2}}{2\pi }%
\;T\sum_{i\omega _{n}}  \notag \\
& \;\Gamma ^{C}(\theta _{1}-\theta _{2})\Upsilon (\theta _{2},q,i\omega
_{n})_{T}\Gamma ^{C}(\theta _{2}-\theta _{1})\Upsilon (\theta _{1},q,i\omega
_{n})_{T}\,,  \label{eq:angularintegration}
\end{align}%
where $\Upsilon (\Delta )_{T}=\big[\Psi (\omega +\Delta )_{T}+\Psi (\omega
-\Delta )_{T}\big]/2$, and $\Gamma ^{C}(\theta )=\nu \Gamma (\theta )$ is
dimensionless angle-dependent amplitude in the Cooper channel. Let us pass
to the Fourier harmonics of the Cooper channel amplitudes, $\Gamma
^{C}(\theta )=\sum_{n}e^{in\theta }\Gamma _{n}^{C}$. We assume that the
amplitudes $\Gamma (\theta )$ are real and are even functions of $\theta
_{1}-\theta _{2}$. This corresponds to real $\Gamma ^{l}=\Gamma ^{-l}$. The
angular integration leads to the following expression for the contribution
of two Cooper sections to the anomalous part of the thermodynamic potential 
\begin{align}
& \Omega _{2}(\Delta )=-\frac{1}{2}\left( \frac{\pi }{2}\right) ^{2}\int 
\frac{qdq}{2\pi }T\sum_{i\omega _{n}}  \notag \\
& \;\Gamma _{l}^{C}\Upsilon _{l-m}(q,i\omega _{n})_{T}\Gamma
_{m}^{C}\Upsilon _{m-l}(q,i\omega _{n})_{T}\,.
\end{align}%
Here the angular harmonic $\Psi _{n}$ of function $\Psi (\theta )$ are
defined as 
\begin{equation}
\Psi _{n}=\int \frac{d\theta }{2\pi }e^{in\theta }\Psi (\theta )\,.
\end{equation}%
After transforming the frequency sum to the integral we obtain 
\begin{align}
& \Omega _{2}(\Delta )=-\sum_{n,m}\frac{\Gamma _{n}^{C}\Gamma _{m}^{C}}{32}%
\int {qdq}\int {d\omega }\coth \frac{\beta \omega }{2}  \notag \\
& \;\Im \big\{\Upsilon _{n-m}(q,\omega ,\Delta )_{T}\Upsilon _{m-n}(q,\omega
,\Delta )_{T}\big\}\,,
\end{align}%
where the contour of $\omega $-integration is slightly shifted above the
real axes.

Let us now turn to the calculation of the anomalous temperature term in the
spin susceptibility. For the chemical potential shifts induced by the
magnetic field $\Delta =2h$ and, therefore, $\delta \chi =-\frac{\partial
^{2}}{\partial {h}^{2}}\Omega _{2}(2h)$. This yields 
\begin{align}
\delta \chi =& \frac{1}{4\pi v_{F}^{2}}\int {d\omega }\coth \frac{\beta
\omega }{2}\Gamma _{n}^{C}\Gamma _{m}^{C}\left\langle h_{nm}(\omega
;s_{1},s_{2})\right\rangle _{T}  \notag \\
h_{nm}(\omega )=& \frac{\pi }{2}\int {qdq}\Im \Big\{\Big[\frac{\partial ^{2}%
}{\partial \omega ^{2}}\Psi _{n-m}(q,\omega _{1})\Big]\Psi _{m-n}(q,\omega
_{2})\Big\}  \notag \\
\phantom{=}& +(\omega _{1}\leftrightarrow \omega _{2})\,.
\label{eq:kappaanom}
\end{align}%
From now on, $q$ is an energy variable, i.e., $qv_{F}\rightarrow {q}$. This
substitution leads to the factor $1/v_{F}^{2}=\pi \nu /\epsilon _{F}$ in the
first line and makes $h_{nm}$ to be dimensionless. The notation $\langle
\cdots \rangle _{T}$ means the integration over $s_{1,2}$ which originates
from Eq.~(\ref{eq:fermi-smearing-Psi}). In $h_{nm}$ the shifted frequencies $%
\omega _{1,2}=\omega -s_{1,2}$ have been introduced and here (and everywhere
below) $\Psi (\omega )$ means $\Psi (\omega )_{T=0}$. Next, in Eq.~(\ref%
{eq:kappaanom}) we replace the second derivative with respect to $\Delta $
by the second derivative with respect to $\omega $ and afterwards the limit $%
h\rightarrow 0$ has been taken. Notice that terms with the first derivatives
vanish because $\partial \Upsilon (h\rightarrow 0)/\partial {h}=0$. The $q$%
-integrations in $h_{nm}$ yields (for details see Appendix~\ref{sec:hnm}) 
\begin{equation}
h_{nm}=-2\mathrm{sign}(\omega _{1}+\omega _{2})\,(-1)^{n-m}\,.
\end{equation}%
It is now necessary to perform the $\omega $ and $s_{1,2}$ integrations 
\begin{align}
& \delta \chi =-\frac{\Gamma _{n}^{C}\Gamma _{m}^{C}(-1)^{n+m}}{2\pi {v}%
_{F}^{2}}\int \frac{ds_{1}ds_{2}}{4}{d\omega }  \notag \\
& \coth \frac{\beta {\omega }}{2}\frac{d\tanh \frac{\beta {s}_{1}}{4}}{ds_{1}%
}\frac{d\tanh \frac{\beta {s}_{2}}{4}}{ds_{2}}\;\mathrm{sign}(2\omega
-s_{1}-s_{2})\,.  \label{eq:chianom}
\end{align}%
The $\omega $-integration can easily be done with the use of the relation $%
\coth (\beta \omega /2)=2Td\ln \sinh (\beta \omega /2)/d\omega $.
Transferring the frequency derivative from $\ln \sinh (\beta \omega /2)$ to $%
\mathrm{sign}(2\omega -s_{1}-s_{2})$ we get a $\delta $-function of
frequency. The boundary term here corresponds to the first integral in the
right hand side of Eq.~(\ref{eq:ultravioletone}). It does not depend on
frequency and will be dropped. Passing to $u_{1,2}=\tanh {\beta {s}_{1,2}}$
as our new integration variables, we get an elementary integral. Finally, 
\begin{equation}
\delta \chi =\nu \frac{T}{\epsilon _{F}}\sum \Gamma _{n}^{C}\Gamma
_{m}^{C}(-1)^{n+m}\,.  \label{eq:resultchianom}
\end{equation}%
The angular integration Eq.~(\ref{eq:angularintegration}) covers all
momentum directions in both Cooper sections. Nevertheless, the result given
in Eq.~(\ref{eq:resultchianom}) can be written in terms of the backward
scattering amplitude only 
\begin{equation}
\delta \chi =\nu \frac{T}{\epsilon _{F}}|\Gamma ^{C}(\pi )|^{2}\,,
\label{eq:secondorderresult}
\end{equation}%
where $\Gamma ^{C}(\pi )=\sum_{n}(-1)^{n}\Gamma _{n}^{C}$. It is crucial for
this result that, remarkably, the $q$-integration in $h_{nm}$ depends on
harmonic indices only through the factor $(-1)^{n-m}$. This also implies
that the scattering amplitude $\Gamma (0)$ drops out from the anomalous
temperature corrections to the spin susceptibility calculated to the second
order in $\Gamma _{l}^{C}$, see Ref.~\onlinecite{Chubukov2003}. The result
obtained here coincides with Eq.~(\ref{eq:linearTcorrection}), which has
been calculated within the $2p_{F}$ channel.

\subsection{Cooper ladder in the thermodynamic potential.}

The anomalous terms in the spin susceptibility generated in the Cooper
channel are described by the following expression 
\begin{equation}
\delta {\chi }=\int \frac{qdq}{2\pi }\;{d\omega }\coth \frac{\beta \omega }{2%
}\;\;\mathrm{Tr}\;\Im \Big\{\partial _{\omega }^{2}\widehat{\Psi }(q,\omega
)_{T}\widehat{\gamma }_{C}\Big\}\,,  \label{eq:ladder}
\end{equation}%
where 
\begin{equation}
\widehat{\gamma }_{C}=\widehat{\Gamma }_{C}\big[1+(\pi /2)\widehat{\Psi }%
(q,\omega )_{T}\widehat{\Gamma }_{C}\big]^{-1}\,.  \label{eq:cooper}
\end{equation}%
Here $(\hat{\Psi})_{nm}=\Psi _{n-m}$ represents a matrix of harmonics of $%
\Psi (\theta )$ and $(\widehat{\Gamma }_{C})_{nm}=\Gamma _{C}^{n}\delta
_{nm} $ is a matrix of the Cooper channel amplitudes. The trace is over
angular harmonic index $n$. Notice that only the terms with the second
derivative with respect to magnetic field acting on the same Cooper section
survive the limit $h\rightarrow 0$.

Let us first explain why the term $\partial _{\omega }^{2}\Psi _{0}(\omega )$
alone does not contribute to $\delta \chi $ (Obviously, such term cannot
exist for non-zero harmonics). Using relation $(\omega \partial
_{q}+q\partial _{\omega })\partial _{\omega }\Psi _{0}(q,\omega )=0$ one can
reduce the $q$-integration to the boundary terms which however vanish
because of absence of the imaginary part. Therefore, at least one more $\Psi 
$ is necessary to be a partner of $\partial _{\omega }^{2}\Psi _{0}(q,\omega
)$. After choosing the partner in the matrix $\widehat{\gamma }_{C}$ the
expression in Eq.~(\pageref{eq:ladder}) takes the form: 
\begin{align}
& \delta {\chi }=\int \frac{qdq}{2\pi }\;{d\omega }\coth \frac{\beta \omega 
}{2}\;\;\;  \notag \\
& \times \mathrm{Tr}\Im \Big\{\partial _{\omega }^{2}\widehat{\Psi }%
(q,\omega )_{T}\widehat{\gamma }_{C}\widehat{\Psi }(q,\omega )_{T}\widehat{%
\gamma }_{C}\Big\}\,.  \label{eq:partner2}
\end{align}%
Now let us point out the special role of the zero harmonic in $\widehat{%
\gamma }_{C}$. Any $(\widehat{\gamma }_{C})_{nm}$ can be constructed
starting from $\Gamma _{n}^{C}\Psi _{n-l^{\prime }}\cdots \Psi _{l^{\prime
\prime }-m}\Gamma _{m}^{C}$ which is chosen to be irreducible with respect
to $\Psi _{0}$. Each $\Gamma _{n}^{C}$ can be dressed by arbitrary number of 
$\Psi _{0}$ by replacing $\Gamma _{n}^{C}$ with $\gamma _{n}=\Gamma
_{n}^{C}/(1+(\pi /2)\Gamma _{n}^{C}\Psi _{0})$. The specifics of $\Psi _{0}$
is that it depends logarithmically on the ultraviolet cutoff. Let us single
out this term, i.e., $\Psi _{0}=\ln (\Lambda /{T})+\widetilde{\Psi }_{0}$.
As we know from the experience of the second order calculation the anomalous
contribution accumulates from $\omega ,qv_{F}\sim {T}$. Therefore $\tilde{%
\Psi}_{0}\sim 1$ as well as $\Psi _{n\neq {m}}$. Now neglecting $\tilde{\Psi}%
_{0}$ and $\Psi _{n\neq {m}}$ in $\widehat{\gamma }_{C}$ we come to the
expression for the anomalous term in the spin-susceptibility, which is
renormalized by Cooper logarithms 
\begin{equation}
\delta \chi =\nu \frac{T}{\epsilon _{F}}\Big|\sum_{n}(-1)^{n}{\gamma }_{n}({T%
})\Big|^{2}\,,  \label{eq:ttilde}
\end{equation}%
where the renormalized $\gamma _{n}({T})$ is 
\begin{equation}
{\gamma }_{n}({T})=\frac{\Gamma _{n}^{C}}{1+\Gamma _{n}^{C}\ln \Lambda /{T}}%
\,.
\end{equation}%
This result is in full correspondence with the renormalization group
equation for the angle-dependent amplitude $\Gamma ^{C}(\theta )$ in the
Cooper-channel 
\begin{equation}
\frac{d\Gamma ^{C}(\theta )}{d\xi }=-\int \frac{d\theta ^{\prime }}{2\pi }%
\Gamma ^{C}(\theta -\theta ^{\prime })\Gamma ^{C}(\theta ^{\prime })\,,
\label{eq:anglescooper}
\end{equation}%
where $\xi =\ln \Lambda /T$. [Notice that apart from renormalizations of $%
\Gamma _{n}^{C}$, the large logarithmic part cannot contribute when a
partner of $\partial _{\omega }^{2}\Psi $ in Eq.~(\ref{eq:partner2}) is
chosen to be the zero harmonic, because of the same argument as presented in
the beginning of this paragraph.]

The result given in Eq.~(\ref{eq:ttilde}) can be presented in terms of the
renormalized backward scattering amplitude 
\begin{equation}
\delta \chi =\nu \frac{T}{\epsilon _{F}}|\Gamma ^{C}(\pi )_{T}|^{2}\,,
\end{equation}%
where $\Gamma ^{C}(\pi )_{T}=\sum_{n}(-1)^{n}{\gamma }_{n}(T)$. For a
general analysis of the temperature dependence of $\delta \chi (T)$ it is
more appropriate to work with $\partial \delta \chi /\partial {T}$ rather
than with the spin susceptibility itself. Unlike the spin susceptibility,
the integrals which determine this quantity converge at $\omega \sim T$ and
depend on the ultraviolet cutoff only through the renormalizations (compare
with Eq.~(\ref{eq:ultravioletone})). The differentiation of the Eq.~(\ref%
{eq:ttilde}) gives 
\begin{equation}
d\delta \chi /d{T}=(\nu /\epsilon _{F})\sum_{nm}(-1)^{n+m}\left[ \gamma
_{n}(T)\gamma _{m}(T)+2\gamma _{n}(T)^{2}\gamma _{m}(T)\right]
\label{eq:mere}
\end{equation}

In fact, the situation is more complicated. The derivative $\partial \chi
/\partial T$ is a series in ${\gamma }_{l}({T})$ which does not reduce to
Eq.~(\ref{eq:mere}): 
\begin{align}
& d\delta \chi /d{T}=(\nu /\epsilon _{F})\sum \Big[(-1)^{l^{\prime
}+l^{\prime \prime }}\gamma _{l^{\prime }}({T})\gamma _{l^{\prime \prime }}({%
T})  \notag \\
& +c_{3}^{l^{\prime }l^{\prime \prime }l^{\prime \prime \prime }}\gamma
_{l^{\prime }}({T})\gamma _{l^{\prime \prime }}({T})\gamma _{l^{\prime
\prime \prime }}({T})+\cdots \Big]\,.  \label{eq:notrg}
\end{align}%
This series is generated when in Eq.~(\ref{eq:ladder}) an arbitrary number
of $\Psi _{n\neq {m}}$ or $\widetilde{\Psi }_{0}$ are taken as a partner for 
$\partial _{\omega }^{2}\Psi $, and it differs from Eq.~(\ref{eq:mere}). The
fact that they are different is obvious when all $l^{\prime }l^{\prime
\prime }\cdots $ are not equal to each other, but this is also valid when
some (or all) of $l$'s are equal. The reason why the series in Eq.~(\ref%
{eq:notrg}) is more sophisticated than a mere differentiation of Eq.~(\ref%
{eq:ttilde}) is because the functions $\Psi _{n\neq {m}}$ and $\widetilde{%
\Psi }_{0}$ depend strongly on the ratio $\omega /qv_{F}$. In the
microscopic Fermi liquid theory\cite{Pitaevskii} the scattering amplitude in
the zero sound channel has exactly the same feature.

Let us point out to a rather subtle contribution to $c_{3}$ which gives
additional evidence that a simple renorm-group generalization of the second
order result given by Eq.~(\ref{eq:ttilde}) is not complete. As it was
indicated in Appendix~\ref{sec:2pf}, the calculation of the product of four
Green's functions contains the term with the integration which is limited by
the ultraviolet cutoff. This term did not contain any temperature dependence
and has been dropped. However, when matrix elements of $\widehat{\gamma }%
_{C}(\omega )$ acquire frequency dispersion (see Eq.~(\ref{eq:cooper})) the
result of this integration ceases to be a featureless constant and it
produces a new contribution to $c_{3}$, which does not reduce to the
renormalization group generalization of the second order result Eq.~(\ref%
{eq:ttilde}).

\section{Calculation of the function $h_{nm}(\protect\omega_1,\protect\omega%
_2)$.}

\label{sec:hnm}

The function $\Psi _{0}(\omega )$ is rather special because it contains the
logarithm of the ultraviolet cutoff. Therefore, we will consider the
contributions involving zero and non-zero harmonics separately. We show that
the result of the $q$-integration in $h_{nm}$ depends on harmonic indices
only through the factor $(-1)^{n-m}$. This remarkable feature is responsible
for the fact that the linear in temperature term in the spin susceptibility
calculated from two Cooper sections depends only on the backward scattering
amplitude $\Gamma ^{C}(\pi )=\sum (-1)^{n}\Gamma _{n}^{C}$.

\subsection{Terms with zero harmonic of $\Psi (\protect\theta )$.}

To obtain an expression for zero angular harmonic $\Psi _{0}$ we first
calculate its imaginary part at frequencies shifted slightly above the real
axis. At zero temperature we get from Eq.~(\ref{eq:Lehman-Psi}) for $\omega
=\omega +i\delta $: 
\begin{align}
\Im \Psi _{0}(q,\omega )& =\mathrm{sign}\omega ,\qquad & |\omega |& >q 
\notag \\
& =(2/\pi )\arcsin (\omega /q),\qquad & |\omega |& <q
\end{align}%
The corresponding analytic expression for $\Psi _{0}$ applicable in the
complex-$\omega $ plane is 
\begin{align}
& \Psi _{0}(q,\omega )  \notag \\
& \;\;=-\frac{1}{\pi }\Big\{-i\pi +{\ln }^{(\rightarrow )}\frac{\omega
+(\omega ^{2}-q^{2})^{1/2}}{\Lambda }  \notag \\
& \qquad \qquad \qquad +{\ln }^{({\leftarrow })}\frac{\omega +(\omega
^{2}-q^{2})^{1/2}}{\Lambda }\Big\}  \label{eq:psizero}
\end{align}%
where $\Lambda $ is an ultraviolet cut-off. The notation ${\ln }%
^{(\rightarrow )}(z)$ and ${\ln }^{({\leftarrow })}(z)$ signifies a
particular choice of the complex logarithmic function. On the real axis the
function ${\ln }^{({\leftarrow })}(z)$ has a branch cut at $z<0$ and it is
real at $z>0$. The function ${\ln }^{(\rightarrow )}$ has a branch cut on
the real axis at $z>0$; its imaginary part is equal to zero on the upper
side of the branch cut and to $2\pi {i}$ on the lower side of the branch
cut. Altogether, near the real axis of $\omega $ the function $\Im \Psi
_{0}=\pm \mathrm{sign}\omega $ for $\omega =\omega \pm \delta $ when $%
|\omega |>q$. The square root $\sqrt{\omega ^{2}-q^{2}}$ in the complex-$%
\omega $ plane has a branch cut between $q$ and $-q$ on the real axes; it is
real and positive at $\omega >q$ and negative at $\omega <-q$.

Let us consider the diagonal part $h_{nn}$. 
\begin{align}
& h_{nn}(\omega) = \frac{\pi}2 \int{qdq} \Im\Big\{\Big[\partial_{\omega}^2%
\Psi_{0}(q,\omega_1)\Big] \Psi_{0}(q,\omega_2)\Big\}  \notag \\
&\qquad +(\omega_1 \leftrightarrow \omega_2)  \notag \\
=& -\frac{\pi}2 \int{dq}\omega_1\frac{\partial}{\partial{q}} \Im\Big\{\Big[%
\partial_{\omega}\Psi_{0}(q,\omega_1)\Big]\Psi_{0}(q,\omega_2)\Big\}  \notag
\\
\phantom{=}& +\frac{\pi}2 \int{dq} \Im\Big\{\,\Big[\partial_{\omega}%
\Psi_{0}(q,\omega_1)\Big]\,\omega_1\frac{\partial}{\partial{q}}%
\Psi_{0}(q,\omega_2) \Big\}  \notag \\
\phantom{=}& + (\omega_1 \leftrightarrow \omega_2) \,.  \label{eq:diagonalh}
\end{align}
To eliminate the second derivative of $\Psi_0(\omega_1)$ we used the
relation $(q\partial_{\omega}+\omega\partial_q)\big[\partial_{\omega}\Psi_0%
\big]=0$ for $\big[\partial_{\omega}\Psi_0\big] = c/(\omega^2-q^2)^{1/2}$ to
replace the $\omega$ derivative with the $q$-derivative. From now on, we use 
$c=-2/\pi$. In the right-hand side of Eq.~(\ref{eq:diagonalh}) the first
term is constructed to be a full derivative.

Now, using the relation $(\omega\partial_{\omega}+q\partial_q)\Psi_0=c$
which follows from the explicit form of $\Psi_0$, we replace in the last
term on the right-hand side of Eq.~(\ref{eq:diagonalh}) the $q$-derivative
with the $\omega$-derivative. The result is 
\begin{align}
&h_{nn}(\omega) = -\frac{\pi}2 \int{\omega_1 dq}\frac{\partial}{\partial{q}}
\Im\Big\{\Big[\partial_{\omega}\Psi_{0}(q,\omega_1)\Big]\Psi_{0}(q,\omega_2)%
\Big\}  \notag \\
\phantom{=}& +\frac{\pi}2 \int{dq} \Im\Big\{\Big[\partial_{\omega}%
\Psi_{0}(q,\omega_1)\Big]\Big( \frac{c\omega_1}q \Big) \Big\}  \notag \\
\phantom{=}& + \frac{\pi}2 \int{dq} \Im\Big\{\Big[\partial_{\omega}%
\Psi_{0}(q,\omega_1)\Big] \left(-\frac{\omega_1\omega_2}q\right) \Big[%
\partial_{\omega}\Psi_{0}(q,\omega_2)\Big]\Big\}  \notag \\
\phantom{=}& + (\omega_1 \leftrightarrow \omega_2) \,.
\end{align}
The integral of the full derivative in the first term is non-zero at the
lower limit. Having in mind $\omega[\partial_{\omega}\Psi_{0}(q\rightarrow0,%
\omega)]=c$ and $\Im\Psi(q\rightarrow0,\omega) = \mathrm{sign}\omega$ we
obtain for it 
\begin{align}
&-\frac{\pi}2 \int{\omega_1 dq}\frac{\partial}{\partial{q}} \Im\Big\{\Big[%
\partial_{\omega}\Psi_{0}(q,\omega_1)\Big]\Psi_{0}(q,\omega_2)\Big\}  \notag
\\
=& -\mathrm{sign}\omega_2 \,.
\end{align}
To calculate the integral in the second line we consider a function $%
c/(\omega^2-q^2)^{1/2}$ in the complex-$q$ plane. For positive $\omega$ we
define the square root in the complex-$q$ plane as a function which has a
branch-cut outside $[-\omega,\omega]$ and has a negative imaginary part
above a brunch-cut at $q>\omega$. With this, we can relate this expression
to the integral in the complex-$q$ plane along the contour $C$ which
envelopes the branch-cut in the counter-clockwise fashion. The integral is
calculated by deforming the contour so that it encloses the pole at $q=0$.
For positive $\omega_1$ it is equal to $-1$; for $\omega_1<0$ it changes the
sign. Overall, we have 
\begin{align}
& \frac{\pi}2\int{dq} \Im\Big\{\Big[\partial_{\omega}\Psi_{0}(q,\omega_1)%
\Big]\Big( \frac{c\omega_1}q \Big)\Big\}  \notag \\
=& -\frac{\pi}{2} \left(\omega_1\Big[\partial_{\omega}\Psi_{0}(q,\omega_1)%
\Big]\right)_{q=0} = -\mathrm{sign}\omega_1 \,.
\end{align}

The integral in the last line of Eq.~(\ref{eq:diagonalh}) can be calculated
by similar method. A delicate point here is to notice all the changes of
signs 
\begin{align}
\phantom{=}& \frac{\pi}2 \int{dq} \Im\Big\{\Big[\partial_{\omega}\Psi_{0}(q,%
\omega_1)\Big] \left(-\frac{\omega_1\omega_2}q\right) \Big[%
\partial_{\omega}\Psi_{0}(q,\omega_2)\Big]\Big\}  \notag \\
&\qquad = \mathrm{sign}(\omega_1+\omega_2)\mathrm{sign}(\omega_1)\mathrm{sign%
}(\omega_2) \,.
\end{align}
Collecting all terms, we find for $h_{nn}$ 
\begin{align}
h_{nn} =& 2\big[\mathrm{sign}(\omega_1+\omega_2) \mathrm{sign}(\omega_1)%
\mathrm{sign}(\omega_2)- \mathrm{sign}\omega_1 - \mathrm{sign}\omega_2] 
\notag \\
=& -2\mathrm{sign}(\omega_1+\omega_2)
\end{align}

\subsection{Terms involving non-zero harmonics of $\Psi(\protect\theta)$.}

To obtain a zero-temperature expression for the non-zero angular harmonics
of $\Psi (\theta )$ we first evaluate the angular integral $\Im \Psi _{n}(q,{%
\omega })=\int (d\theta /2\pi )\mathrm{sign}({\omega }-qv_{F}\cos \theta
)e^{in\theta }$ 
\begin{align}
\Im \Psi _{n}(q,{\omega })& =0,\;\; & |{\omega }|& >q  \notag \\
& =-(2/n\pi )\sin \Big(n\arccos (\omega /q)\Big),\;\; & |{\omega }|& <q\,.
\end{align}%
In the complex-$\omega $ plane the analytic expression for $\Psi _{n}$ has a
form 
\begin{equation}
\Psi _{n}(\omega )=\frac{2}{\pi |n|}\left( \frac{\omega -(\omega
^{2}-q^{2})^{1/2}}{q}\right) ^{|n|}\,.  \label{eq:nondiagonal}
\end{equation}%
We need to calculate the $q$-integral in ${h}_{nm}(\omega )=\tilde{h}%
_{n-m}(\omega _{1},\omega _{2})+\tilde{h}_{n-m}(\omega _{2},\omega _{1})$
where 
\begin{equation}
\tilde{h}_{l}(\omega _{1},\omega _{2})=\frac{\pi }{2}\int {qdq}\Im \Big\{%
\Big[\frac{\partial ^{2}}{\partial \omega ^{2}}\Psi _{l}(q,\omega _{1})\Big]%
\Psi _{l}(q,\omega _{2})\Big\}\,.  \label{eq:delicate}
\end{equation}%
Let us sketch the way we perform the $q$-integration in this case. When both 
$\omega _{1}$ and $\omega _{2}$ are positive, the $q$-integral can be done
using contours in the complex-$q$ plane. It yields $\tilde{h}_{l}(\omega
_{1}>0,\omega _{2}>0)=-(-1)^{l}$. Let us look on this result in more
details. The integral can be written as $(\pi /2)\int {qdq}[\Im A\Re B+\Re
A\Im B]$, where $A=\partial _{\omega }^{2}\Psi _{l}(q,\omega _{1})$ and $%
B=\Psi _{l}(q,\omega _{2})$. When $\omega _{1}<\omega _{2}$ the lower limit
for the term with $\Re A\Im B$ is $q=\omega _{2}$. The integral $(\pi
/2)\int_{\omega _{2}}^{\infty }{qdq}(\Re A\Im B)$ is purely trigonometric.
It is identified as $-1$ for even $l$ and it is zero otherwise.
Correspondingly, the integral $(\pi /2)\int_{\omega _{1}}^{\infty }{qdq}(\Im
A\Re B)$ is $1$ for odd $l$ and zero otherwise. Alternatively, for $\omega
_{2}<\omega _{1}$ the integral $(\pi /2)\int_{\omega _{1}}^{\infty }{qdq}%
(\Im A\Re B)$ is purely trigonometric. It is identified as $-1$ for even $l$
and zero otherwise. We now use this information to resolve the delicate
question of the sign changes in various domains in $\omega _{1}$, $\omega
_{2}$ plane.

A peculiar feature of the $q$-integration is that at any value of $\omega $
we take the value of $\Psi $ on the upper side of its branch cut in the
complex-$\omega $ plane. From the definition of $\sqrt{\omega ^{2}-q^{2}}$
given after Eq.~(\ref{eq:psizero}), it follows that for $\omega $ slightly
above the real axis $\Psi _{l}(-\omega )=(-1)^{l}\Psi _{l}^{\ast }(\omega )$%
; the same holds for $A_{l}(\omega )$ and $B_{l}(\omega )$. We start with
the case when $\omega _{1}<0$ and $|\omega _{1}|<\omega _{2}$. In this case
the $q$-integral in $\tilde{h}_{l}(\omega _{1},\omega _{2})$ has to be
defined as $(\pi /2)\int {qdq}(-1)^{l}[-\Im A\Re B+\Re A\Im B]$. Using the
values for the two $q$-integrals found above, we obtain for this combination
again $-(-1)^{l}$. Therefore, the $q$-integral in $\tilde{h}_{l}(\omega
_{1},\omega _{2})$ does not change when $\omega _{1}$ changes its sign. When 
$|\omega _{2}|<\omega _{1}$ and $\omega _{2}<0$ the $q$-integral in $\tilde{h%
}_{l}(\omega _{1},\omega _{2})$ becomes $(\pi /2)\int {qdq}(-1)^{l}[\Im A\Re
B-\Re A\Im B]$. Here again, this combination reproduces $-(-1)^{l}$. Hence,
the integral remains constant when it is extended across the boundaries of
the quadrant $\omega _{1},\omega _{2}>0$. On the other hand, when both $%
\omega _{1}$ and $\omega _{2}$ are negative the integral has opposite sign, $%
\tilde{h}_{l}(\omega _{1}<0,\omega _{2}<0)=(-1)^{l}$. Altogether, it follows
from this analysis that $\tilde{h}_{l}(\omega _{1},\omega _{2})$ changes
sign when $\omega _{1}+\omega _{2}=0$; i.e., 
\begin{equation}
h_{nm}=-2\,\mathrm{sign}({\omega _{1}+\omega _{2}})(-1)^{n-m}\,.
\label{eq:h-nm}
\end{equation}

\begin{acknowledgements}
We thank E. Abrahams, A.V.~Chubukov, K.B.~Efetov, P.A.~Lee, E.~Mizrahi, C.P.~Nave, A.~Punnoose and G.~Schwiete for valuable
discussions. We are grateful to M.~Reznikov for providing us with the detailed information about the
experimental situation. A.F. is supported by the Minerva Foundation and by the Rosa and Emilio Segra Research Award.
A.S. is supported by the Maurice Goldschleger Center for Nanophysics  and by I2CAM fellowship, grant NSF DMR 0645461. We thank the Department of
Material Sciences in Argonne National Laboratory for hospitality; the work was supported by U.S. Department of
Energy under Contract No. W-31-109-ENG-38.
\end{acknowledgements}


\begin{thebibliography}{99}
\bibitem{zerosound} For the amplitude $\Gamma(\pi)$ there is also a
possibility to look on it in the zero-sound channel. Here a delicate
question arises whether a singular combination $v_{F}k/(\omega-v_{F}k)$
typical for the particle-hole channel\cite{Pitaevskii} is of any importance.
The calculation in the Appendix~\ref{sec:twoGreen} shows that the anomalous
part in the product of four Green's functions accumulates from $v_{F}k\sim {v}_{F}p_{\perp}\gg\omega$. Therefore, when it is treated in the Cooper channel,  the bare value for the amplitude $\Gamma(\pi)$ is the static amplitude $\Gamma ^{k}$ in the particle-hole
channel.
\bibitem{twodiff} Notice that this statement relies on the fact that in the spin susceptibility there are only two differentiations.
\bibitem{threesections}  It has been checked that these terms involve combinations of harmonics which do not reduce to the backward scattering amplitude. Unlike the two section term, there is only a weak logarithmic singularity near the backward scattering, which is not enough to make the logarithmic renormalizations important in channels other than the Cooper channel.
\end{thebibliography}
\end{document}